\documentclass[copyright]{eptcs}
\providecommand{\event}{EXPRESS 2012}

\usepackage{breakurl}
\usepackage{version}
\usepackage{amssymb,amsmath,amscd,amsfonts,latexsym}
\usepackage{tikz}
\usepackage{pgf}
\usetikzlibrary{arrows,automata,backgrounds,decorations}
\usepgflibrary{decorations.pathreplacing}
\usepackage{graphicx}
\usepackage{xypic}

\RequirePackage[T1]{fontenc}
\RequirePackage{times}

\newcommand{\trans}[1][]{\stackrel{#1}{\longrightarrow}}
\newcommand{\emptypr}{\varepsilon}
\newcommand{\eps}{\varepsilon}

\newcommand{\thread}{thread}
\renewcommand{\a}{\alpha}

\newcommand{\smallsection}[1]{\vspace{0.2cm}\noindent{\bf #1.} }

\newcounter{examplecounter}
\newenvironment{myexample}{
  \refstepcounter{examplecounter}
  \bigskip
  \noindent{\textbf{Example \arabic{examplecounter}.}}
}

\newcommand{\len}[1]{|#1|}

\newcommand{\removed}[1]{}
\newenvironment{proofof}[1]{\smallskip \noindent{\bf Proof of #1.}}{\hfill $\Box$} 
\newcommand{\slcomm}[1]{}
\newcommand{\bpppump}{{\sc \CCFL-pumping}}
\newcommand{\strongpump}{{\sc concat. pumping}}
\newcommand{\weakpump}{{\sc shuffle pumping}}
\newcommand{\malyprefix}[1]{$_{^{\mathrm{#1}}}$}
\newcommand{\malyprefixpodw}[2]{$^{_{\mathrm{#1}}}_{^{\mathrm{#2}}}$}

\newcommand{\BPC}{{\sc BPC}}
\newcommand{\CFL}{{\sc cfl}}
\newcommand{\CCFL}{\malyprefix{c}\CFL}  
\newcommand{\PCCFL}{\malyprefix{pc}\CFL}  
\newcommand{\tPCCFL}{\malyprefixpodw{tr}{pc}\CFL}
\newcommand{\PAL}{\malyprefix{shuffle}\CFL}
\newcommand{\PA}{{\sc pa}}
\newcommand{\TCFL}{\malyprefix{trace}\CFL}
\newcommand{\parcomp}{\, || \,}
\newcommand{\seqcomp}{ ; }
\newcommand{\nrof}[2]{\text{\tiny \#}_{#1}(#2)}

\newtheorem{proposition}{Proposition}
\newtheorem{theorem}{Theorem}
\newtheorem{lemma}{Lemma}
\newtheorem{remark}{Remark}

\title{Partially-commutative context-free languages}

\author{Wojciech Czerwi{\'n}ski
\institute{Institute of Informatics, University of Warsaw
\thanks{The first author acknowledges a partial support by the Polish MNiSW grant N N206 568640.}}
\email{wczerwin@mimuw.edu.pl}
\and
S{\l}awomir Lasota
\institute{Institute of Informatics, University of Warsaw
\thanks{The second author acknowledges a partial support by the Polish MNiSW grant N N206 356036.}}
\email{sl@mimuw.edu.pl}
}

\def\titlerunning{Partially-commutative context-free languages}
\def\authorrunning{W. Czerwi{\'n}ski \& S. Lasota}

\begin{document}

\maketitle

\begin{abstract}
The paper is about a class of languages that extends context-free languages (\CFL) and is stable under shuffle.
Specifically, we investigate the class of \emph{partially-com\-mu\-ta\-tive context-free languages} (\PCCFL),
where non-ter\-mi\-nal symbols are commutative according to a binary independence relation, 
very much like in trace theory. 
The class has been recently proposed as a robust class subsuming \CFL\  and
commutative \CFL.
This paper surveys properties of \PCCFL.
We identify a natural corresponding automaton model: stateless multi-pushdown automata. 
We show stability of the class under natural operations, including homomorphic images and shuffle. 
Finally, we relate expressiveness of \PCCFL\ to two other relevant classes: \CFL\ extended with shuffle  
and trace-closures of \CFL.
Among technical contributions of the paper are pumping lemmas, as an elegant completion of known pumping properties of regular languages, \CFL\ and commutative \CFL.
\end{abstract}

\section{Introduction}


Closure of languages classes under shuffle is intensively investigated, see for instance~\cite{BBCPR10} 
and further references therein.
This paper is about a subtle way of introducing shuffle into context-free grammars.

\smallsection{Process algebraic motivation} 
In the context of infinite-state verification there are two basic well known classes of systems.
\emph{Context-free processes}, called traditionally BPA\footnote{A shorthand for Basic Process Algebra.}~\cite{BK84}, stand for the most fundamental abstract model of sequential recursive programs.
BPA contains configuration graphs induced by context-free grammars in Grei\-bach normal form.
The commutative variant, \emph{commutative context-free processes}, traditionally called BPP\footnote{A shorthand for Basic Parallel Processes Algebra.}, was proposed in~\cite{PHDChristensen} as the abstract model of concurrent programs. 
BPP differs from BPA in that it has parallel composition instead of sequential composition. 
Thus a configuration is a finite multiset of non-terminals rather than a sequence.

A natural generalization of both BPA and BPP is \emph{Process Algebra} (\PA)~\cite{BK84} where one allows for both kinds of composition.\footnote{The algebra of~\cite{BK84} includes also left merge operation, not considered in this paper.}
A standard reference for a process-rewrite formulation of PA is~\cite{Mayr00}. 
However, \PA\ does not seem to have good algorithmic properties. For instance, bisimulation equivalence is not known to be decidable, a long standing open problem~\cite{Srba02}, while the algorithm for normed PA is very complex and as costly as double exponential time~\cite{DBLP:conf/icalp/HirshfeldJ99}.
This has recently motivated investigation of an alternative but equally natural generalization of both BPA and BPP,
namely \emph{partially-commutative context-free processes}, called \BPC\footnote{A shorthand for Basic Partially Commutative Algebra.} in~\cite{CFL09}.
\BPC\ processes are also defined by a Greibach grammar, but one additionally assumes a binary 
\emph{independence relation} among non-terminals, like in trace theory~\cite{Mazurkiewicz88}, 
and only independent pairs of non-terminals commute.
We stress that the independence is imposed not on alphabet letters, which is usually the case in trace languages, but on non-terminals. Thus a configuration may be modeled as a trace over non-terminals. 
BPA is a special case where no non-terminals commute while BPP, on the other hand, is another special case where all non-terminals commute. 

In~\cite{CFL09,CFL11}  an efficient polynomial-time procedure has been developed for bisimulation equivalence, that works correctly in the subclass of normed \BPC\ that strictly contains both normed BPA and BPP.
We also very recently analyzed the reachability problem for \BPC~\cite{CHL12}.
In this paper we continue the program that aims at finding a robust class subsuming BPA and BPP,
however this time from the language-theoretic perspective.

\smallsection{Language theoretic motivation}
BPA clearly defines context-free languages (\CFL) and BPP defines so called commutative context-free languages (\CCFL)~\cite{PHDChristensen}\footnote{In fact, BPA and BPP define \CFL\ and \CCFL\ not containing the empty word,
respectively.}, equivalently characterized as languages of com\-mu\-ni\-ca\-tion-free Petri nets. 
In this paper we focus on \emph{partially-commutative context-free languages} (\PCCFL)~\cite{CFL11} that are defined by \BPC. Our aim is to investigate properties of this class and to relate its expressiveness with other classes.

The class \PCCFL\ extends \CFL\ and is closed under the shuffle operation.
By a shuffle of two words we mean here an arbitrary interleaving of these words and
by shuffle of two languages we mean all shuffles of all pairs of words from the two languages.
Other similar extensions of \CFL\ may be found in the literature.
One such extension is \PA\ languages. We use a shorthand \PAL\ for this class --
as far as languages are concerned, \PA\ is equivalent to context-free grammars where one allows 
to use both concatenation and shuffle in productions~\cite{Gischer81,NSS03}.
Another related class is trace-closures of \CFL\ (name this class \TCFL), 
where one assumes, contrary to \PCCFL, an independence relation on alphabet letters. 
We have found it appealing to relate the expressive power of \PCCFL\ with \PAL\ and \TCFL.
 
\smallsection{Our contribution}
First, we show that a relevant subclass of \PCCFL, subject to the restriction that the complement of independence relation is transitive, has a natural corresponding automaton model: stateless multi-pushdown automata (Section~\ref{sec:prelim}). We also prove that the membership problem for \PCCFL\ is NP-complete thus the complexity remains the same as for \CCFL\
 (Section~\ref{sec:derivtrees}).

Second, in Section~\ref{sec:closure}  we investigate stability of \PCCFL\ under natural operations. In particular, \PCCFL\ turns out to be stable under homomorphic images, substitutions and shuffle. On the other hand, the class is not stable under inverse homomorphic images and under intersections with regular languages. The latter is not very surprising as we consider a natural extension of \CCFL, 
the class that lacks not only the two closure properties, but even lacks closure under concatenation and homomorphic images! With \PCCFL\ one regains closure under concatenation and homomorphic images.

Third, in Sections~\ref{sec:plemmas} and~\ref{sec:express}  we perform mutual comparison of expressiveness of \PCCFL, transitive \PCCFL, \PAL\ and \TCFL, proving them all pairwise incomparable (except for the trivial inclusion of transitive \PCCFL\ in \PCCFL, that we prove to be strict). 
Note that incomparability with respect to languages implies incomparability with respect to bisimulation or other equivalences.
As one of the tools we formulate and prove pumping lemmas for classes \PCCFL, transitive \PCCFL\ and \PAL.
This provides an elegant completion of known pumping properties of regular, context-free and commutative context-free languages.

Technically, the most difficult part is Sections~\ref{sec:plemmas} and~\ref{sec:express}. On the other hand,
the results of Sections~\ref{sec:prelim} and~\ref{sec:closure} confirm clearly
that \PCCFL \ is a natural class of languages extending \CFL, with good algorithmic and closure properties.


Yet another relevant language class is that defined by so called Dynamic Pushdown Networks~\cite{BMT05}.
The class extends \CFL\ and is closed under shuffle. We do not investigate this class here,
but we conjecture that it is incomparable with \PCCFL.

Some of the proofs are omitted 
due to space limitation.

\section{Preliminaries}\label{sec:prelim}

By an \emph{interleaving} of two words $w$ and $v$, of length $m$ and $n$, respectively,
 we mean any word $u$ of length $m+n$ such that its positions $I = \{1, \ldots, m+n\}$ may be
split into two disjoint sets $I_w$ and $I_v$ such that $u$ restricted to $I_w$ equals $w$
and $u$ restricted to $I_v$ equals $v$. Let $w \parcomp v$ denote the set of all the interleavings
of $w$ and $v$, which is clearly a finite set.
By a shuffle of two languages $L$ and $K$ we mean
\[
L \parcomp K  =  \bigcup_{w \in L, v \in K} w \parcomp v .
\]

\smallsection{Partially-commutative context-free languages}
\label{sec:pccfl}
The class of languages to be defined below has been introduced in~\cite{CFL11}, 
however our presentation and terminology here is different.

A Greibach context-free grammar consists of a finite alphabet, a finite set of non-terminal symbols $V$ with a distinguished initial symbol $S\in V$ and a finite set of productions of the form
\begin{equation}
\label{eq:gnf-rule}
X \trans[a] \alpha,
\end{equation}
where $X \in V$, $\alpha \in V^*$ and $a$ is an alphabet letter. Additionally we assume that a grammar is always equipped with a symmetric and irreflexive 
relation $I\subseteq V\times V$ called the \emph{independence relation}.
For convenience we also use the complement $D = (V\times V) \setminus I$, called the \emph{dependence relation}.
Two non-terminals $X, Y\in V$ are called \emph{independent} if $(X,Y)\in I$, and otherwise \emph{dependent}.

Any $\alpha \in V^*$ we call a \emph{configuration}.
A \emph{derivation} is a sequence of configurations 
such that every configuration is obtained from the preceding one via a \emph{step} and the last one is the empty configuration. There are two kinds of steps:
\begin{itemize} 
\item \emph{production step}: $X \beta \trans[a] \alpha \beta$, for a production $X \trans[a] \alpha$;
\item \emph{swap step}: $\alpha X Y \beta \trans \alpha Y X \beta$, where $X$ and $Y$ are independent.
\end{itemize}
Every derivation \emph{defines} a word $w$ obtained by concatenation of alphabet letters occurring in the production steps.
We write $\alpha \trans[w] \beta$ if there is a derivation that defines $w$, starts in $\alpha$ and ends in $\beta$.
We usually assume that a derivation starts with a configuration consisting of a single non-terminal, say $X$.
If $X \trans[w] \emptypr$ then we say that $X$ \emph{generates} $w$.
Note that the length of $w$ is the same as the number of production steps performed in any derivation that defines $w$. We assume wlog.~that every non-terminal $X$ generates some word.

The language generated by a grammar is the set of all words generated by the initial non-terminal.
The class of all so generated languages we call 
\emph{partially-commutative context-free languages} (\PCCFL)~\cite{CFL11}.
It clearly contains all context-free languages (\CFL) and commutative 
context-free languages\footnote{The commutative context-free languages are also called \emph{BPP languages}.} (\CCFL)~\cite{PHDChristensen}. 
These two subclasses are special cases, where independence is either the identity, or the full relation, respectively.

\begin{myexample}\label{ex:nottransitive}
For illustration, consider the grammar:
\[
\begin{array}{cclccclccclcccl}
P & \trans[a] & W B C \bar{B} &\hspace{0.5cm}& 
W & \trans[a] & W B C   &\hspace{0.5cm}& 
\bar{B} & \trans[\bar{b}] & \emptypr &\hspace{0.5cm}& B & \trans[b] & \emptypr \\
&&&& 
W & \trans[\bar{a}] & \bar{C} &&       
\bar{C} & \trans[\bar{c}] & \emptypr && C & \trans[c] & \emptypr 
\end{array}
\]

The initial non-terminal is $P$ and the independence relation is the symmetric closure of $\{B, \bar{B}\} \times \{C, \bar{C}\}$.
Here is an example derivation of the word $a \bar{a} b \bar{b} \bar{c} c$.
\[
\begin{array}{rcccccccccc}
P & \trans[a] & W B C \bar{B} & \trans[\bar{a}] &
\bar{C} B C \bar{B} & \trans &
\bar{C} B \bar{B} C  & \trans &
B \bar{C} \bar{B} C  & \trans \\
&& B \bar{B}  \bar{C} C &   
\trans[b] & \bar{B} \bar{C} C & \trans[\bar{b}] & \bar{C} C & \trans[\bar{c}] & C & \trans[c] & \eps 
\end{array}
\]
In a similar way a word $a^n \bar{a} b^n \bar{b} \bar{c} c^n$ is generated, for any $n \geq 1$, 
but also $a^n \bar{a} \bar{c} c^n b^n \bar{b}$ or $a^n \bar{a} \bar{c} b^n c^n \bar{b}$.
The language generated is
\[
\bigcup_{n \geq 1} a^n \bar{a} \, (b^n \bar{b} \parcomp \bar{c} c^n).
\]
\end{myexample}

We might have defined configurations as Ma\-zur\-kie\-wicz traces~\cite{Mazurkiewicz88} rather than words over non-terminals
(like in~\cite{CFL11}).
This would mean that \emph{trace equivalent} configurations are not distinguished. 
In our terminology, two configurations are trace equivalent when one may be transformed into another using solely swap steps.
It is our deliberate choice to keep the swap steps explicit.

\removed{ 
\smallsection{Silent productions}
In the above definitions one could allow for \emph{silent production}, i.e., with $a = \varepsilon$.
Clearly, the resulting class \PCCFL$_{\varepsilon}$ is strictly larger, 
as languages in \PCCFL\ never contain the empty word $\varepsilon$.
We claim that the extension would be non-trivial:
\begin{proposition}\label{p:epsPCCFL}
\PCCFL$_{\varepsilon}$ $\setminus$ \PCCFL\ contains a language $L$ with $\varepsilon \notin L$.
\end{proposition}
Having said that, we claim that most of our results remain valid if silent productions are allowed.
In this paper we do not consider silent productions for simplicity.
} 

\smallsection{Transitive dependence}
We distinguish a subclass of \PCCFL\ where dependence is assumed to be transitive, being thus an equivalence. 
This subclass we name \tPCCFL.
Equivalence classes of dependence will be called \emph{{\thread}s}. 

In Example~\ref{ex:nottransitive} the dependence is not transitive, as it contains $(P,B)$ and $(P,C)$ but not $(B,C)$.
In fact we show later that this language does not belong to \tPCCFL.
Both \CFL\ and \CCFL\ are strict subclasses of \tPCCFL.

\begin{myexample}\label{ex:transitive}
As an illustration, consider the language generated by:
\[
\begin{array}{ccccccccccccccc}
S & \trans[s] & \emptypr &\hspace{0.5cm}& 
S & \trans[a] & S A  &\hspace{0.5cm}& 
A & \trans[c] & A' &\hspace{0.5cm}& 
A' & \trans[a] & \emptypr \\
&&&&       
S & \trans[b] & S B  && 
B & \trans[c] & B' && 
B' & \trans[b] & \emptypr 
\end{array}
\]

with initial non-terminal $S$ and the {\thread}s $\{S, A, B\}$, $\{A'\}$ and $\{B'\}$.
Here is an example derivation  of the word $absccab$.
\[
S\trans[a] S A \trans[b] S B A \trans[s] B A \trans[c] B' A \trans A B'
\trans[c] A' B' \trans[a] B' \trans[b] \eps.
\]
The language contains words of the form $w s v$, 
where $w$ contains only $a$ and $b$ and $v$ contains only $a$, $b$ and $c$.
Writing $\nrof{a}{w}$ for the number of occurrences of $a$ in $w$ and $|w|$ for the length of $w$,
we may characterize the language by the following conditions:
\begin{itemize}
\item $\nrof{a}{w} = \nrof{a}{v}$, $\nrof{b}{w} = \nrof{b}{v}$ and $\nrof{c}{v} = \nrof{a}{v} + \nrof{b}{v}$,
\item any prefix $v'$ of $v$ and any suffix $w'$ of $w$ such that $\nrof{c}{v'} = |w'|$ fulfills
\[
    \nrof{a}{w'} \geq \nrof{a}{v'} \qquad \text{and} \qquad
    \nrof{b}{w'} \geq \nrof{b}{v'}.
\]
\end{itemize}
\end{myexample}

\smallsection{Automaton model}
A multi-pushdown automaton is like a single-pushdown one.
In a single step one symbol is popped from one of the stacks,\footnote{If we allowed for popping from more than one stack at a time,
the model would clearly become Turing-complete, even with only one state.} and a number of symbols are
pushed on the stacks. 
The number of stacks is fixed for an automaton.
Assume there is only one state, or equivalently no state, and $k$ stacks.
Then a transition of an automaton is of the form:
\begin{equation} \label{eq:mpdatran}
X \trans[a] \a_1 \ldots \a_k,
\end{equation}
to mean that when an automaton reads $a$, it pops $X$ and pushes the sequence of symbols $\a_i$ on the $i$th stack,
 for $i = 1 \ldots k$. Observe that wlog. one may assume that stack alphabets are disjoint. The following result is an easy observation:
\begin{theorem}[\cite{CHL12}]
The \tPCCFL\ class is expressively equivalent to stateless multi-push\-down automata.
\end{theorem}
Indeed, an equivalence class of configurations with respect to trace equivalence is represented
by a tuple of strings, one per \thread. Similarly, a production $X \trans[a] \alpha$ is represented, up to swap steps, 
exactly as in~\eqref{eq:mpdatran}, with $\a_i$ being the projection of $\a$ on the $i$th \thread.

Similarly, one could also define an operational model for general \PCCFL, with a stack replaced by a partially ordered structure.

\section{Derivation trees} \label{sec:derivtrees}

It is very convenient to use derivation trees instead of derivations themselves.
However it is not completely obvious how to define this notion in presence of commutativity of non-terminals.
Below we adopt an intuitive approach using colors.

Fix a derivation $X \trans[w] \emptypr$.
Clearly a configuration is a sequence of \emph{non-terminal occurrences}.
We assume that every non-terminal occurrence in a derivation will be colored, including the occurrence of $X$ in the initial configuration.
We impose the following simple discipline of coloring:
\begin{itemize}
  \item if a swap step $\alpha X Y \beta \trans \alpha Y X \beta$ is performed, every non-terminal occurrence in the right-hand 
  side configuration
  inherits its color from the corresponding occurrence of the same non-terminal on the left-hand side.
  \item if a production step $X \beta \trans[a] \alpha \beta$ is performed, the non-terminal occurrences in $\beta$ preserve 
  their colors, while all the non-terminals occurrences in $\alpha$ get fresh colors. Note that the color of the occurrence of $X$ 
  in the beginning of $X \beta$ disappears as a result of the step. We say that this disappearing color \emph{drops} the fresh colors. 
\end{itemize}
Intuitively, a color is intended to represent the 'life cycle' of one occurrence of a non-terminal during a derivation.
Observe that non-terminal occurrences in a given configuration are always labeled with different colors, and that the total
number of colors used in a derivation equals the number of production steps.

\begin{myexample}
A disciplined coloring of the derivation from Example~\ref{ex:transitive} is shown below.
Colors are $1, 2, \ldots$ and the coloring is denoted by subscripts. 
\begin{equation} \label{eq:deriv}
S_1\trans[a] S_2 A_3 \trans[b] S_4 B_5 A_3 \trans[s] B_5 A_3 \trans[c] B'_6 A_3 \trans A_3 B'_6
\trans[c] A'_7 B'_6 \trans[a] B'_6 \trans[b] \eps.
\end{equation}


Color $1$ drops colors $2$ and $3$, color $3$ drops color $7$, etc.
\end{myexample}

With the use of our coloring discipline, every derivation induces naturally a tree.
The tree nodes are all colors appearing in the derivation. The color $c_1$ is a parent of $c_2$ precisely if $c_1$ drops $c_2$.
Every tree node $c$ is labeled by a non-terminal. If convenient, one may think that
every node is labeled by a production that made color $c$ disappear.

There may be many different derivations inducing the same tree.
Even worse, two derivations \emph{of different words} may induce the same tree, as shown in the example below.

\begin{myexample} \label{ex:derivtree}
Continuing the last example, the derivation~\eqref{eq:deriv} induces the following tree:
\begin{center}
\begin{tikzpicture}[inner sep=3pt, level distance=1cm, sibling distance=5cm]
  \node at (0,0) {$1 : {S \trans[a] S A}$}
  child[sibling distance=3cm] {node {$2 : {S \trans[b] S B}$}
    child {node{$4 : {S \trans[s] \eps}$}}
    child {node{$5 : {B \trans[c] B'}$}
      child {node{$6 : {B' \trans[b] \eps}$}}}}
  child {node {$3 : {A \trans[c] A'}$}
    child {node {$7 : {A' \trans[a] \eps}$}}};
\end{tikzpicture}
\end{center}
However, exactly the same tree is induced by the derivation:
\[
S_1 \trans[a] S_2 A_3 \trans[b] S_4 B_5 A_3 \trans[s] B_5 A_3 \trans[c] B'_6 A_3 \trans[b] A_3 \trans[c] A'_7 \trans[a] \eps
\]
of a different word $abscbca \neq absccab$. Intuitively, the words defined by subtrees rooted in $3$ and $6$, namely $c a$ and $b$ respectively, this time come in a different order. In fact all the interleavings of these two words are allowed.
\end{myexample}

\newcommand{\propis}{{\sc Induced subword}}
\newcommand{\propir}{{\sc  Infix rearrangement}}
\newcommand{\props}{{\sc Substitutivity}}
\smallsection{Useful properties}
The examples confirm that our notion of derivation tree is more complex than the classical one.
However, trees may be still very useful for reasoning about partially-commutative context-free languages,
as they immediately bring to light the following useful properties:
\vskip 1mm
\noindent{\propis.}
Given a derivation tree of a word $w$, every node $c$ induces a subword 
(i.e. a subsequence but not an infix in general) of $w$.
Indeed, the subword is obtained by concatenating only those letters from $w$
whose color, as a tree node, belongs to the subtree rooted in $c$.
We implicitly assign here to the letter of every production step a color that disappears in this step.
For instance, for both words considered in the last example, the subword induced by the node $2$ is $bscb$.
Analogously one defines the subword induced by a subset of nodes of a derivation tree, assuming this subset
 to be an antichain with respect to the tree ancestor relation.
\vskip 1mm
\noindent{\propir.}
The induced subword may be rearranged into an infix. Let $L \in $ \PCCFL\ and
let $v$ be the subword of $w \in L$ induced by a tree node $c$.
Clearly, $w \in v \parcomp u$,
i.e., $v$ is interleaved with the remaining subword $u$ of $w$.
Then $u$ may be split into $u = u_1 u_2$ so that $u_1 v \, u_2 \in L$. 
Indeed, let $u_1$ be the prefix of $w$ preceding the first letter of $v$.
In any derivation, after $u_1$, the non-terminal that labels $c$ is clearly active.
Performing the whole derivation $X \trans[v] \emptypr$ immediately after $u_1$ does the job.
\vskip 1mm
\noindent{\props.}
In any derivation tree, one may replace a subtree rooted in a node $c$ 
by an arbitrary derivation tree $t$, 
assumed that both $c$ and the root of $t$ are labeled with the same non-terminal. The resulting tree is 
clearly induced by some derivation too.

\smallsection{Membership problem}
A derivation tree is of linear size in terms of the length of the word, which is useful for easily obtaining 
the upper bound for the membership problem, where
given a word $w$ and a presentation of a language $L$, one asks if $w \in L$?
\begin{theorem}
The membership problem is NP-complete both for \PCCFL\ and \tPCCFL.
\end{theorem}
NP-hardness follows easily from NP-hardness of the membership problem for
\CCFL, shown in~\cite{DBLP:journals/fuin/Esparza97}. The NP upper bound
one obtains easily: guess a tree and the order of its nodes, and then check in polynomial time
whether the tree is induced by some derivation of the given word that respects the order of nodes.

\section{Closure properties}\label{sec:closure}

In this section we argue that  \PCCFL\ and \tPCCFL\ classes are closed under union and shuffle, and
 \PCCFL\ is closed under concatenation while \tPCCFL\ is not.
Then we show that \PCCFL\ is closed under homomorphic images and substitutions. 
In case of \tPCCFL\ we do not know the answer, however we suppose it is negative.
Finally, we show that both classes lack closure under inverse homomorphic images and
intersections with regular languages.

Comparing \PCCFL\ with \CFL, roughly speaking, one sacrifices intersection with regular languages and inverse homomorphic images but one gains shuffle.
Even if at first sight the properties listed above do not seem exciting, one should remember that 
both the classes considered here subsume also commutative context-free languages \CCFL. 
Knowing that \CCFL\ lacks closure under concatenation and homomorphic images, as shown in~\cite{PHDChristensen},
it seems that with \PCCFL\ one \emph{retrieves} these relevant closure properties.
This seems to confirm that \PCCFL\ is a natural class of languages.

\smallsection{Union and complement}
Both classes are closed under union and the construction is entirely standard.
On the other hand none of the classes is closed under complement.
\removed{ 
Assume two languages $L_1$ and $L_2$ belonging either to \tPCCFL\ or to \PCCFL.
Let $S_1$ and $S_2$ be the initial non-terminals of the corresponding grammars.
Then the grammar for $L_1 \cup L_2$ is straightforwardly obtained by adding to the union of two grammars
one additional initial non-terminal $S$ and additional productions
$S \trans[a] \alpha$ for all productions $S_1 \trans[a] \alpha$ or $S_2 \trans[a] \alpha$.
The dependence relation is the set-theoretic union of the two dependence relations,
thus transitivity of dependence is clearly preserved.
} 

\smallsection{Shuffle and concatenation}
Both classes are closed under shuffle and the construction of a grammar for the shuffle 
$L_1 \, || \, L_2$ is easy. 
Wlog assume that the grammars that generate the two languages use distinct non-terminals.
Let $S_1$ and $S_2$ be the initial non-terminals.
Consider the union of grammars extended with one additional initial non-terminal $S$.
Add additional productions
\begin{equation} \label{eq:shuffleinit}
S \trans[a_1] \alpha_1 S_2  \qquad S \trans[a_2] \alpha_2 S_1
\end{equation}
for any production $S_1\trans[a_1]\alpha_1$ or $S_2\trans[a_2]\alpha_2$.
Finally, extend independence by imposing that whenever two non-terminals come from
different grammars they are independent. This clearly preserves transitivity of dependence.

In \PCCFL, concatenation $L_1 L_2$ is obtained similarly as shuffle.
The only difference is that two non-terminals coming from different grammars are always declared dependent,
and that only the left-hand productions in~\eqref{eq:shuffleinit} are added.
Note that concatenation is in our setting no more natural than shuffle.

\tPCCFL\ is not closed under concatenation, which one shows similarly as for \CCFL~\cite{PHDChristensen}.
Consider $L_1 = \{ w : \nrof{a}{w} = \nrof{b}{w} = \nrof{c}{w} \geq 1, \nrof{d}{w} = 0 \}$ and $L_2 = \{ d \}$.
In the derivation of some $w \in L_1 L_2$ a configuration is necessarily reached with at least two different {\thread}s nonempty,
as otherwise the language would be context-free. Thus the remaining suffix of $w$ is some shuffle of at least two words generated by 
these non-empty {\thread}s, and only one of these words ends with $d$. If that subword is generated first, the whole word is not in $L_1 L_2$, which proves that $L_1 L_2$ may not belong to \CCFL.

\smallsection{Homomorphic images and substitutions}
As we consider only Greibach grammars, the empty word never belongs to a partially-commutative context-free language.
Thus it is natural to consider only homomorphisms $h$ that do not contain the empty word in the image: $h(a) \neq \varepsilon$ for all letters $a$. 
Below we show that \PCCFL\ is closed under images of such homomorphisms. For \tPCCFL\ the question is still open;
we conjecture however a negative answer. 

We prefer to show a slightly stronger result: \PCCFL\ is closed under \emph{substitutions}.
A substitution $s$ assigns to each alphabet letter $a$ a language $s(a) \in $ \PCCFL. Similarly as above, we assume that the languages
$s(a)$ do not contain the empty word. For a language $L$, the substitution $L[s]$ contains all words that may be obtained from a word in $L$, by replacing each letter $a$ with any word from $s(a)$.

Assume a language $L \in$ \PCCFL, generated by a grammar $G$, and a substitution $s$. Thus each language $s(a)$ has its generating grammar $G_a$. We describe the construction of the grammar $G'$ for $L[s]$. The non-terminals of $G'$ will be the union of non-terminals of $G$ and all grammars $G_a$. Wlog we assume that the non-terminal sets are disjoint.

Consider an arbitrary production
$X \trans[a] \alpha$ in $G$.
Let $S_a$ be the initial non-terminal in $G_a$. For any production
$S_a \trans[b] \beta$
in $G_a$, we add to $G'$ the production:
$X \trans[b] \beta \alpha$.
The independence in $G'$ is defined as the set-theoretic union of independence relations of grammars $G$ and $G_a$.
Thus any pair of non-terminals coming from different grammars is declared dependent (note that this is not achievable if the dependence has to be transitive).

The construction guarantees that $G'$ generates exactly $L[s]$. Indeed, once a production $X \trans[b] \beta \alpha$ is fired, the non-terminals of $G_a$ block activity of other non-terminals, due to the dependence, until a word of $s(a)$ is generated.

We do not know whether the \tPCCFL\ class is closed under homomorphic images; however we suppose it is not. We conjecture that a counterexample is given by the language
\[
L = \{w: \nrof{a}{w}  = \nrof{b}{w} = \nrof{c}{w}, \nrof{d}{w} = 1\}
\]
together with the homomorphism $h(a) = a$, $h(b) = b$, $h(c) = c$, $h(d) = dd$.

\smallsection{Intersection with regular languages}
Both classes \PCCFL\ and \tPCCFL\ lack closure under intersection with regular languages.
Let $L=\{w: \nrof{a}{w} = \nrof{b}{w} = \nrof{c}{w} \}$. Clearly $L \in $ \CCFL\ but
$L \cap a^*b^*c^*$ is not in \PCCFL\ (and also not in \PAL\ defined in a moment) according to:
\begin{lemma}\label{anbncn}
The language $L=\{a^n b^n c^n: n \geq 1\}$ is not in \PCCFL\ $\cup$ \PAL.
\end{lemma}
It is worth noting that the lack of closure is not surprising as the emptiness problem
for intersection of a partially-com\-mu\-ta\-tive context-free language with a regular language is undecidable,
even if the dependence is assumed to be transitive.
Roughly speaking \tPCCFL\ correspond to stateless multi-pushdown
automata and intersection with regular language corresponds do adding the state which makes the
model Turing powerful.

\smallsection{Inverse homomorphic images}
Both  \PCCFL\ and \tPCCFL\ are not closed under inverse homomorphic images.
Consider the shuffle $L = L_1 \, || \, L_2$  of two context-free languages 
\[
L_1=\{A^{n+1} S B^n T : n \geq 1 \} \qquad L_2=\{S B^n T C^n : n \geq 1 \},
\]
and  
the homomorphism $h$ given by $h(a)=A$, $h(s)=SS$, $h(b)=BB$, $h(t)=TT$ and $h(c)=C$. 
If $h^{-1}(L) = \{a^{n+1} s b^n t c^n: n \geq 1\}$ were in \PCCFL\ then its image under a homomorphism $g(s) = b, g(t) = c$,
that is the language $L$ in Lemma~\ref{anbncn}, would be in \PCCFL\ as well -- a contradiction.

\removed{ 
\smallsection{Complement}

None of the classes is closed under complement as otherwise one would deduce closure under intersection, using the law
\begin{equation*}
L\cap K = \overline{\overline{L}\cup\overline{K}} .
\end{equation*}
} 

\section{Other extensions of context-free languages}

There are two other language classes know from the literature that,
similarly as \PCCFL, extend \CFL\ with some amount of commutation.

\smallsection{\PA\ languages}
The formalism to be described below is traditionally called \emph{Process Algebra} (\PA)~\cite{BK84,Mayr00}.
It is however nothing else than an extension of Greibach context-free grammars
with an explicit shuffle operation: a production has the form
\[
X \trans[a] t,
\]
where $t$ is an arbitrary term built from non-terminals using binary operations of sequential composition '$\seqcomp$' and
parallel composition '$\parcomp$'. The first operation one may interpret as concatenation of languages, and the second one as shuffle (thus the overloading of the symbol $\parcomp$ is absolutely deliberate). 
The \emph{empty term} $\emptypr$ is also allowed.

For convenience, terms are only considered up to a structural equivalence, that imposes associativity of both operations, commutativity of $\parcomp$, and neutrality of $\emptypr$ with respect to
both operations.

A \emph{configuration} is an arbitrary term of the above form.
Steps between configurations are defined by the following rules (the last rule is in fact redundant due to commutativity of $\parcomp$, but we prefer to keep it for readability):
\[
\frac{X \trans[a] t \text{ is a production}}{X \trans[a] t}  
\qquad
\frac{t \trans[a] t'}{t \seqcomp u \trans[a] t' \seqcomp u}
\qquad
\frac{t \trans[a] t'}{t \parcomp u \trans[a] t' \parcomp u}
\qquad
\frac{u \trans[a] u'}{t \parcomp u \trans[a] t \parcomp u'}
\]
As usual, a derivation is a sequence of configurations starting from a distinguished initial configuration $S$, ending in the
empty configuration, such that every subsequent configuration is obtained from a preceding one by a single step.
Other notions, including the language generated by a grammar, or derivation trees, may be defined similarly as for \PCCFL.
The class of languages we denote by \PAL.

In particular, \PAL\ satisfy the three properties mentioned above: \propis, \propir\ and \props.

The difference between \PCCFL\ and \PAL\ is, roughly, a difference between
specifying commutation explicitly in productions, or implicitly by an independence relation.

\smallsection{Trace-closures of \CFL}
To define \TCFL\ we need to assume that an independence
relation ranges not over non-terminals but over alphabet letters instead.
As usual, one defines \emph{trace equivalence} over words:
two words are equivalent if one may be transformed into another by swaps of neighboring independent letters.
A context-free language $L$ is not closed under this equivalence in general and  its \emph{trace closure}
\[
\{w : w \text{ is trace equivalent to some }  v \in L \}
\]
is in general not context-free. By \TCFL\ we denote the class containing trace closures of context-free languages. 
Clearly \TCFL\ is a superclass of \CFL.

\section{Pumping lemmas}\label{sec:plemmas}

Now we analyze how much the classical idea of pumping extends from \CFL\ to larger classes.
Roughly speaking, the intuitive cutting and pasting in a derivation tree does not 
translate to the property of a language as easily as in the case of \CFL.

We formulate two different pumping lemmas. 
Remarkably, with one of them we complete nicely the picture of pumping lemmas known 
for regular, context-free and commutative context-free languages.

As expected, the pumping lemmas appear to be useful tool for relating the expressive power of language classes, as we demonstrate in Section~\ref{sec:express}.

\smallsection{The pumping lemmas}
The length of a word $w$ is written $\len{w}$. To motivate our conditions we start by recalling the 
pumping scheme proposed for \CCFL\ 
by~\cite{PHDChristensen}.

\begin{quote}
(\bpppump~\cite{PHDChristensen})\ \ 
\emph{There is a constant $N$ such that if $w\in L$ with $\len{w} > N$ then there exist words $x, y, s$ such that
\begin{enumerate}
\item $w \in x\, (s \parcomp y)$,
\item $1 \leq \len{s} \leq N$, and
\item $\forall m\geq 0$,  $x\, s^m y \in L$.\footnote{In fact in~\cite{PHDChristensen}, 
the pumping scheme was $x\, s^m y'$, with a suffix $y'$ of $w$ (think of $y' \in s \parcomp y$), rather than  $x\, s^m y$. The proofs of both are very similar. 
We discuss this issue further in Remark~\ref{rem:pumpscheme}.}
\end{enumerate}
}
\end{quote}

\noindent
Point 1 reads as: $w$ is a concatenation of some prefix $x$ and an interleaving of $s$ and $y$.
We define now two new conditions on a language $L$.   

\begin{quote}
(\weakpump)\ \ 
\emph{There is a constant $N$ such that if $w\in L$ with $\len{w} > N$ then there exist words $x, y, z, s, t$ such that
\begin{enumerate}
\item $w \in x\, ((s\, (y \parcomp t)) \parcomp z)$,
\item $1 \leq \len{s}, \len{s\, y\, t} \leq N$, and
\item $\forall m\geq 0$,  $x\, s^m y\, t^m z \in L$.
\end{enumerate}
}
\end{quote}
Point 1 reads as: there is some subword $y'$ of $w$ with $w \in x\, (y' \parcomp z)$ and $y' \in s\, (y \parcomp t)$.
\begin{quote}
(\strongpump)\ \ 
\emph{There is a constant $N$ such that if $w\in L$ with $\len{w} > N$ then there exist words $x, y, z, s, t$ such that
\begin{enumerate}
\item $w = x\, y\, z$,
\item $1 \leq \len{s\, t} \leq N$, and
\item $\forall m\geq 0$,  $x\, s^m y\, t^m z \in L$.
\end{enumerate}
}
\end{quote}
Call the words $s$, $t$ \emph{repeatable words}.
The difference between the two conditions concentrates on the word $y$ that separates the repeatable words in 
$x\, s^m y\, t^m z$.
On one hand \weakpump\ seems weaker as $y$ is no more an infix of $w$, but an arbitrary subword (subsequence). 
On the other hand \weakpump\ seems stronger as the length of $y$ is bounded.


\begin{lemma}\label{lemma-PCCFL}
Every language $L \in$ \PCCFL\ $\cup$ \PAL\ satisfies \weakpump.
\end{lemma}
As an example of application we provide now a proof missing in Section~\ref{sec:closure}.

\begin{proofof}{Lemma~\ref{anbncn}}
Assume towards contradiction that $L=\{a^n b^n c^n: n\geq 1\}$ is in \PCCFL\ or in \PAL\ and 
apply Lemma~\ref{lemma-PCCFL}.
Observe that the two repeatable words $s$ and $t$ have necessarily jointly the same number of letters $a$, $b$ and $c$.
Thus one of them has to contains two different letters. Repeating this word twice leads to a contradiction.
\end{proofof}
\begin{lemma}\label{lemma-tPCCFL}
Every language $L \in$ \tPCCFL\ $\cup$ \PAL\ satisfies \strongpump.
\end{lemma}
Class \PCCFL\ does not satisfy \strongpump, as witnessed by the language from Example~\ref{ex:nottransitive}.
Moreover in \strongpump\ one can not bound the length of the word $y$.  

\smallsection{Relating conditions}
The condition \weakpump\ is similar to the classical context-free pumping --
the only difference is the words $s$, $y$, $t$ and $z$ are subwords, not necessarily infixes, of $w$.
We claim it is an elegant completion of the pumping lemmas for 
regular languages (RL), context-free languages (\CFL) and commutative context-free languages (\CCFL) (see~\cite{PHDChristensen}). 
All of these lemmas may be characterized by the following two characteristics:
\begin{enumerate}
\item Are there one or two pumping positions?
\item Are repeatable words infixes or subwords a given word?
\end{enumerate}
The known pumping lemmas have the following characteristics:
\begin{itemize}
\item RL: 1 pumping position, a repeatable word is an infix
\item \CFL: 2 pumping positions, repeatable words are infixes
\item \CCFL: 1 pumping position, a repeatable word is a subword~\cite{PHDChristensen}.
\end{itemize}
In this light, our condition \weakpump\ offers an elegant completion of the picture: 2 pumping positions, repeatable words are subwords.
In other words, \weakpump\ weakens \CCFL-pumping in the same way as \CFL-pumping weakens RL-pumping (2 pumping positions instead of one).
The other way around: \weakpump\ weakens \CFL-pumping in the same way as \CCFL-pumping weakens RL-pumping (repeatable word is no more an infix). The relationships between the four pumping conditions is depicted in the following diagram:
\[
\xymatrix @C-4pc{
& \ar@{--}[rrrddd] & \text{\weakpump} &\ar@{--}[lllddd] & \\
\ar@{--}[rrdd] &  \text{\ \ \ \ \ \ \ \CFL-pumping\ \ } \ar@{=>}[ru] && \text{\ \ \CCFL-pumping\ \ \ } \ar@{=>}[lu] & \ar@{--}[lldd] \\
\txt{two pumping\\positions} && \text{RL-pumping} \ar@{=>}[ru] \ar@{=>}[lu] && \txt{repeatable subword} \\
&\txt{\ \ one pumping\\ \ position} && \txt{repeatable infix} &
} 
\]

\begin{remark} \label{rem:pumpscheme}
It is worth mentioning that another pumping scheme could be used in place of \weakpump\ in Lemma~\ref{lemma-PCCFL}: instead of
$
x\, s^m y\, t^m z,
$
one may consider
\[
x\, s^m y'\, t^m z,
\]
with $w \in x\, (y' \parcomp z)$ and $y' \in (s\, (y \parcomp t))$.
The proof would be very similar.
\end{remark}

\section{Expressiveness}\label{sec:express}

Now we are ready to compare the expressive power of \tPCCFL\ and \PCCFL\ with other classes.
We show that \tPCCFL\ is a strict subclass of \PCCFL\ and that
both \PAL\ and \TCFL\ are incomparable with either \PCCFL\ or \tPCCFL.
More specifically, our results are as follows:
\newcommand{\thmstrict}{\tPCCFL\ is a strict subclass of \PCCFL}
\begin{theorem} \label{thm:strict}
\thmstrict.
\end{theorem}
\newcommand{\thmtPCCFL}{\tPCCFL\ is not included in \PAL\ $\cup$ \TCFL}
\newcommand{\thmPAL}{\PAL\ is not included in \PCCFL\ $\cup$ \TCFL}
\newcommand{\thmTCFL}{\TCFL\ is not included in \PCCFL\ $\cup$ \PAL}
\newcommand{\thmtPCCFLcapPAL}{\tPCCFL\ $\cap$ \PAL\ is not included in \TCFL}
\newcommand{\thmtPCCFLcapTCFL}{\tPCCFL\ $\cap$ \TCFL\ is not included in \PAL}
\begin{theorem} \label{thm:non-incl-add}
The following non-inclusions hold:
\begin{enumerate}
\item[(1)] \thmtPCCFLcapPAL.
\item[(2)] \thmtPCCFLcapTCFL.
\end{enumerate}
\end{theorem}

\begin{figure}[h]
\begin{center}
\includegraphics[width = 72mm, height = 37mm]{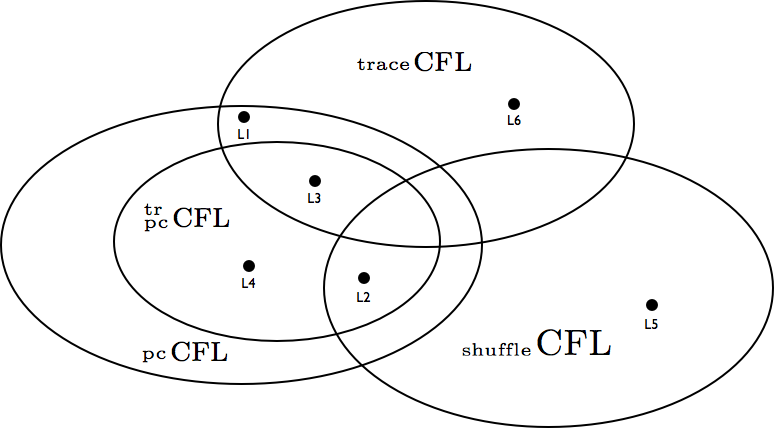}
\end{center}
\caption{Relating the expressive power.}
\label{fig:express}
\end{figure}

\begin{theorem} \label{thm:non-incl}
The following non-inclusions hold:
\begin{enumerate}
\item[(1)] \thmtPCCFL;
\item[(2)] \thmPAL;
\item[(3)] \thmTCFL.
\end{enumerate}
\end{theorem}
The proofs of the results are by identifying witnessing languages
$L_1$ \ldots $L_6$, as illustrated in Figure~\ref{fig:express}.
The pumping lemmas, namely Lemma~\ref{lemma-tPCCFL} and Lemma~\ref{lemma-PCCFL},
are sufficient to prove Theorem~\ref{thm:strict} and Theorem~\ref{thm:non-incl}(3), respectively.
On the other hand they are not sufficient for
Theorem~\ref{thm:non-incl-add}(2) and Theorem~\ref{thm:non-incl}(1)--(2),
as \PAL\ satisfies both the lemmas, and thus we have to perform a more delicate analysis of a derivation tree.
We illustrate the first kind of argument in the proof of Theorem~\ref{thm:non-incl}(3) and the second kind in the
proof of Theorem~\ref{thm:non-incl}(2) below.


\newcommand{\sal}{\bar{a}}
\newcommand{\sdl}{\bar{d}}

\begin{proofof}{Theorem~\ref{thm:non-incl}(3)}
Consider the language
\begin{equation}\label{shuffle_form}
L_6 = \{w \in \bigcup_{n\geq 0} \Big( a^n \sal\, \sdl\, d^n \, || \, b^n c^n \Big)
 : \text{every } b \text{ preceds every } d \text{ and } \sdl \text{ in } w \} .
\end{equation}
Clearly, $L_6$ is the trace closure of the context-free language
$
\{
(a\,b)^n \sal\, \sdl\, (c\,d)^n 
: n\geq 0\},
$
if for the independence on alphabet letters one chooses the symmetric closure of:
\[
\{a, \sal\} \times \{
b, c
\} \;\cup\; \{ \sdl, d \} \times \{ c \} .
\]
Using Lemma~\ref{lemma-PCCFL} we will show that
$L_6$ belongs to neither \PCCFL\ nor \PAL.
Consider a word 
\[
w_n = a^n \sal\, 
b^n c^n 
\sdl\, d^n
\]
and recall that for $n$ larger that $N$ of Lemma~\ref{lemma-PCCFL} we would obtain
\[
w_n \in x\, (y' \parcomp z) \qquad y' \in s\, (y \parcomp t)
\]
for a substring $y'$ of $w_n$. Recall also the pumping scheme of \weakpump\ from Lemma~\ref{lemma-PCCFL}:
\begin{equation} \label{eq:recall-pumping}
x s^m y t^m z \in L_6, \ \ \text{ for } m \geq 0.
\end{equation}
We do a sequence of simple observations. First, to keep the same number of appearances of letters $a, b, c$ and $d$,
each of the four letters must appear either in $s$ or $t$.
Second, both $s$ and $t$ are necessarily non-empty as otherwise we would observe an illegal order of letters in~\eqref{eq:recall-pumping}, and moreover $a$ and $b$ occur in $s$ and $c$ and $d$ occur in $t$,
keeping in mind that in $L_6$ every $a$ precedes every $d$ and every $b$ precedes every $c$ and $d$.
Third, the length of the prefix $x$ is at most $n$, as otherwise both $s$ and $t$ would appear to the right of $\sal$ and thus could not contain $a$.
\slcomm{zmiana dowodu - recenzent 2}
Thus, $x$ contains only $a$. Now, $\overline{d}$ is not in $x$,
cannot be in $s$ or $t$, and cannot be in $z$ since otherwise
($s$ and) $t$ could not contain $d$. Therefore $\overline{d}$
is in $y$, and $z$ contains no $b$. As neither $x$ nor $z$
contains $b$, and $w_n\in x(y'\parallel z)$, $y'$ must contain
$n$ occurrences of $b$, but $|y'|=|syt|\leq N$, hence this is not
possible.
%
%
We have thus shown that $L_6$ does not satisfy Lemma~\ref{lemma-PCCFL} and therefore it does not belong to \PCCFL\ $\cup$ \PAL.
\end{proofof}

\begin{proofof}{Theorem~\ref{thm:non-incl-add}(2)}
Consider the language $L_3 \in $ \tPCCFL:
\begin{equation}\label{eq:languagel3}
L_3 = \bigcup_{n \geq 0} a^n s \, (b^n \parcomp c^n)
\end{equation}
 and a grammar that generates the language:
\[
\begin{array}{ccccccccccc}
S & \trans[a] & SP & \hspace{0.5cm} &
P & \trans[b] & C & \hspace{0.5cm} &
C & \trans[c] & \emptypr \\
S & \trans[s] & \emptypr & \hspace{0.5cm} &
P & \trans[c] & B & \hspace{0.5cm} &
B & \trans[b] & \emptypr
\end{array}
\]

The initial non-terminal is $S$ and the threads are $\{S, P\}, \{B\}, \{C\}$.
$L_3$ also belongs to \TCFL\ as it is the trace closure of the context-free language 
$\{a^n s (bc)^n : n \geq 0\}$ with independence $\{(b,c), (c,b)\}$.

It remains thus to show that $L_3 \notin$ \PAL.
Intuitively, the idea is to show that $L_3$ cannot benefit from parallel composition.

Assume that $L_3 \in $ \PAL, aiming at deducing a contradiction.
Fix a grammar that generates $L_3$. For simplicity think of the productions of the 
following form
(the first two we will call \emph{sequential}):
\[
X \trans[a] \emptypr \qquad
X \trans[\varepsilon] Y \seqcomp Z \qquad
X \trans[\varepsilon] Y \parcomp Z .
\]
We will exploit the property that  $s$ divides every word in $L_3$
into two separated regions.
We partition the non-terminals into symbols that generate some word
containing $s$, and symbols that do not; and call them $s$-symbols and 
non-$s$-symbols, respectively.
By \props, each word generated by an $s$-symbol contains necessarily $s$.

Consider a derivation tree $T$ of a word $w s v \in L_3$.
The unique path leading from the root to the leaf labeled by $s$ call
\emph{the spine}.
Observe that an $s$-symbol may only appear on the spine and
a non-$s$-symbol may only appear outside the spine.
Knowing that the number of occurrences of $a$ and $b$ on both sides of the spine
is the same, we deduce that 
\begin{align} \label{eq:spine-seq}
\text{each production labeling a node of the spine is necessarily sequential.}
\end{align}
Indeed, assume a parallel production $X \trans[\varepsilon] Y \parcomp Z$ 
labels a node of the spine. Wlog. let $Y$ be a $s$-symbol.
Let $u$, $u'$ be the subwords induced by the $Y$-node and $Z$-node, respectively.
Clearly there are two interleavings of $u$ and $u'$  such that
the letter $s$, appearing in $u$, is placed in the interleaving in two different positions in the word $u'$.
Thus at least one of these interleavings must lead to a violation of the condition~\eqref{eq:languagel3}
in a word belonging to $L_3$.
Condition~\eqref{eq:spine-seq} is thus proved.

Now consider a non-$s$-symbol $X$ appearing in $T$.
The number of occurrences $\nrof{a}{u}$ of $a$ in all words $u$ generated by $X$ is necessarily
the same, and the same applies to $\nrof{b}{u}$ and $\nrof{c}{u}$.
Indeed, otherwise one gets a similar contradiction as above by considering two words
induced by the $X$ node, differing in the number of occurrences of $a$ or $b$, and using \props.
As a consequence $X$ generates a finite language which may clearly be
defined by a context-free grammar, say $G_X$. 

If we apply the last observation to the very first non-$s$-symbol $X$ on every path
in $T$ (except the spine), we obtain a tree without parallel nodes.
As $G_X$ does not depend on the particular derivation tree $T$ chosen, and the word $w s v \in L_3$ 
was chosen arbitrary, we conclude that $L_3$ is generated by a context-free grammar.
The grammar is obtained by replacing productions of every non-$s$-symbol $X$ in $G$ with $G_X$.
As $L$ is clearly not context-free we obtain a contradiction and thus complete the proof.
\end{proofof}


\nocite{*}
\bibliographystyle{eptcs}
\bibliography{citat}

\begin{thebibliography}{10}
\providecommand{\bibitemdeclare}[2]{}
\providecommand{\surnamestart}{}
\providecommand{\surnameend}{}
\providecommand{\urlprefix}{Available at }
\providecommand{\url}[1]{\texttt{#1}}
\providecommand{\href}[2]{\texttt{#2}}
\providecommand{\urlalt}[2]{\href{#1}{#2}}
\providecommand{\doi}[1]{doi:\urlalt{http://dx.doi.org/#1}{#1}}
\providecommand{\bibinfo}[2]{#2}

\bibitemdeclare{article}{BK85}
\bibitem{BK85}
\bibinfo{author}{J.~A. \surnamestart Bergstra\surnameend} \&
  \bibinfo{author}{J.~W. \surnamestart Klop\surnameend} (\bibinfo{year}{1985}):
  \emph{\bibinfo{title}{Algebra of Communicating Processes with Abstraction}}.
\newblock {\sl \bibinfo{journal}{Theor. Comput. Sci.}} \bibinfo{volume}{37},
  pp. \bibinfo{pages}{77--121}, \doi{10.1016/0304-3975(85)90088-X}.

\bibitemdeclare{article}{BK84}
\bibitem{BK84}
\bibinfo{author}{Jan~A. \surnamestart Bergstra\surnameend} \&
  \bibinfo{author}{Jan~Willem \surnamestart Klop\surnameend}
  (\bibinfo{year}{1984}): \emph{\bibinfo{title}{Process Algebra for Synchronous
  Communication}}.
\newblock {\sl \bibinfo{journal}{Information and Control}}
  \bibinfo{volume}{60}(\bibinfo{number}{1-3}), pp. \bibinfo{pages}{109--137},
  \doi{10.1016/S0019-9958(84)80025-X}.

\bibitemdeclare{article}{BBCPR10}
\bibitem{BBCPR10}
\bibinfo{author}{Jean \surnamestart Berstel\surnameend}, \bibinfo{author}{Luc
  \surnamestart Boasson\surnameend}, \bibinfo{author}{Olivier \surnamestart
  Carton\surnameend}, \bibinfo{author}{Jean-Eric \surnamestart Pin\surnameend}
  \& \bibinfo{author}{Antonio \surnamestart Restivo\surnameend}
  (\bibinfo{year}{2010}): \emph{\bibinfo{title}{The expressive power of the
  shuffle product}}.
\newblock {\sl \bibinfo{journal}{Inf. Comput.}}
  \bibinfo{volume}{208}(\bibinfo{number}{11}), pp. \bibinfo{pages}{1258--1272},
  \doi{10.1016/j.ic.2010.06.002}.

\bibitemdeclare{inproceedings}{BMT05}
\bibitem{BMT05}
\bibinfo{author}{Ahmed \surnamestart Bouajjani\surnameend},
  \bibinfo{author}{Markus \surnamestart M{\"u}ller-Olm\surnameend} \&
  \bibinfo{author}{Tayssir \surnamestart Touili\surnameend}
  (\bibinfo{year}{2005}): \emph{\bibinfo{title}{Regular Symbolic Analysis of
  Dynamic Networks of Pushdown Systems}}.
\newblock In: {\sl \bibinfo{booktitle}{CONCUR}}, pp. \bibinfo{pages}{473--487},
  \doi{10.1007/11539452\_36}.

\bibitemdeclare{phdthesis}{PHDChristensen}
\bibitem{PHDChristensen}
\bibinfo{author}{S.~\surnamestart Christensen\surnameend}
  (\bibinfo{year}{1993}): \emph{\bibinfo{title}{Decidability and Decomposition
  in Process Algebras}}.
\newblock Ph.D. thesis, \bibinfo{school}{Department of Computer Science,
  University of Edinburgh}.

\bibitemdeclare{inproceedings}{CFL09}
\bibitem{CFL09}
\bibinfo{author}{W.~\surnamestart Czerwi{\'n}ski\surnameend},
  \bibinfo{author}{S.~B. \surnamestart Fr{\"o}schle\surnameend} \&
  \bibinfo{author}{S.~\surnamestart Lasota\surnameend} (\bibinfo{year}{2009}):
  \emph{\bibinfo{title}{Partially-Commutative Context-Free Processes}}.
\newblock In: {\sl \bibinfo{booktitle}{CONCUR}}, pp. \bibinfo{pages}{259--273},
  \doi{10.1007/978-3-642-04081-8\_18}.

\bibitemdeclare{article}{CFL11}
\bibitem{CFL11}
\bibinfo{author}{W.~\surnamestart Czerwi{\'n}ski\surnameend},
  \bibinfo{author}{S.~B. \surnamestart Fr{\"o}schle\surnameend} \&
  \bibinfo{author}{S.~\surnamestart Lasota\surnameend} (\bibinfo{year}{2011}):
  \emph{\bibinfo{title}{Partially-commutative context-free processes:
  expressibility and tractability}}.
\newblock {\sl \bibinfo{journal}{Inf. Comput.}}
  \bibinfo{volume}{209}(\bibinfo{number}{5}), pp. \bibinfo{pages}{782--798},
  \doi{10.1016/j.ic.2010.12.003}.

\bibitemdeclare{unpublished}{CHL12}
\bibitem{CHL12}
\bibinfo{author}{W.~\surnamestart Czerwi{\'n}ski\surnameend},
  \bibinfo{author}{P.~\surnamestart Hofman\surnameend} \&
  \bibinfo{author}{S.~\surnamestart Lasota\surnameend} (\bibinfo{year}{2012}):
  \emph{\bibinfo{title}{Reachability problem for weak multi-pushdown
  automata}}.
\newblock \bibinfo{note}{To appear}.

\bibitemdeclare{article}{DBLP:journals/fuin/Esparza97}
\bibitem{DBLP:journals/fuin/Esparza97}
\bibinfo{author}{J.~\surnamestart Esparza\surnameend} (\bibinfo{year}{1997}):
  \emph{\bibinfo{title}{Petri Nets, Commutative Context-Free Grammars, and
  {Basic Parallel Processes}}}.
\newblock {\sl \bibinfo{journal}{Fundam. Inform.}}
  \bibinfo{volume}{31}(\bibinfo{number}{1}), pp. \bibinfo{pages}{13--25},
  \doi{10.1007/3-540-60249-6\_54}.

\bibitemdeclare{article}{Gischer81}
\bibitem{Gischer81}
\bibinfo{author}{Jay~L. \surnamestart Gischer\surnameend}
  (\bibinfo{year}{1981}): \emph{\bibinfo{title}{Shuffle Languages, Petri Nets,
  and Context-Sensitive Grammars}}.
\newblock {\sl \bibinfo{journal}{Commun. ACM}}
  \bibinfo{volume}{24}(\bibinfo{number}{9}), pp. \bibinfo{pages}{597--605},
  \doi{10.1145/358746.358767}.

\bibitemdeclare{inproceedings}{DBLP:conf/icalp/HirshfeldJ99}
\bibitem{DBLP:conf/icalp/HirshfeldJ99}
\bibinfo{author}{Y.~\surnamestart Hirshfeld\surnameend} \&
  \bibinfo{author}{M.~\surnamestart Jerrum\surnameend} (\bibinfo{year}{1999}):
  \emph{\bibinfo{title}{Bisimulation Equivalence Is Decidable for Normed
  {Process Algebra}}}.
\newblock In: {\sl \bibinfo{booktitle}{ICALP}}, pp. \bibinfo{pages}{412--421},
  \doi{10.1007/3-540-48523-6\_38}.

\bibitemdeclare{article}{Mayr00}
\bibitem{Mayr00}
\bibinfo{author}{R.~\surnamestart Mayr\surnameend} (\bibinfo{year}{2000}):
  \emph{\bibinfo{title}{Process Rewrite Systems}}.
\newblock {\sl \bibinfo{journal}{Inf. Comput.}}
  \bibinfo{volume}{156}(\bibinfo{number}{1-2}), pp. \bibinfo{pages}{264--286},
  \doi{10.1006/inco.1999.2826}.

\bibitemdeclare{inproceedings}{Mazurkiewicz88}
\bibitem{Mazurkiewicz88}
\bibinfo{author}{A.~W. \surnamestart Mazurkiewicz\surnameend}
  (\bibinfo{year}{1988}): \emph{\bibinfo{title}{Basic notions of trace
  theory}}.
\newblock In: {\sl \bibinfo{booktitle}{REX Workshop}}, pp.
  \bibinfo{pages}{285--363}.

\bibitemdeclare{inproceedings}{NSS03}
\bibitem{NSS03}
\bibinfo{author}{M.-J. \surnamestart Nederhof\surnameend},
  \bibinfo{author}{G.~\surnamestart Satta\surnameend} \&
  \bibinfo{author}{S.~\surnamestart Shieber\surnameend} (\bibinfo{year}{2003}):
  \emph{\bibinfo{title}{Partially Ordered Multiset Context-Free Grammars And
  Free-Word-Order Parsing}}.
\newblock In: {\sl \bibinfo{booktitle}{Proc. 8th Intl Workshop on Parsing
  Technologies}}, pp. \bibinfo{pages}{171--182}.

\bibitemdeclare{article}{Srba02}
\bibitem{Srba02}
\bibinfo{author}{J.~\surnamestart Srba\surnameend} (\bibinfo{year}{2002}):
  \emph{\bibinfo{title}{Roadmap of Infinite Results}}.
\newblock {\sl \bibinfo{journal}{Bulletin of the EATCS}} \bibinfo{volume}{78},
  pp. \bibinfo{pages}{163--175}.

\end{thebibliography}

\end{document}